\documentclass[twocolumn,aps]{revtex4-1}

\pdfpageattr {/Group << /S /Transparency /I true /CS /DeviceRGB>>}

\usepackage{graphicx} 
\usepackage{overpic}
\usepackage{enumitem}
\usepackage{amssymb}

\DeclareSymbolFont{lettersA}{U}{txmia}{m}{it}
\DeclareMathSymbol{\real}{\mathord}{lettersA}{"92}
\DeclareMathSymbol{\field}{\mathord}{lettersA}{"83}

\begin{document}

\title{Minimum weight perfect matching of fault-tolerant topological quantum error correction in average $O(1)$ parallel time}

\author{Austin G. Fowler$^{1,2}$}

\affiliation{$^1$Department of Physics, University of California, Santa Barbara, California 93106, USA \\
$^2$Centre for Quantum Computation and Communication Technology, School of Physics, The University of Melbourne, Victoria 3010, Australia}

\date{\today}

\begin{abstract}
Consider a 2-D square array of qubits of extent $L\times L$. We provide a proof that the minimum weight perfect matching problem associated with running a particular class of topological quantum error correction codes on this array can be exactly solved with a 2-D square array of classical computing devices, each of which is nominally associated with a fixed number $N$ of qubits, in constant average time per round of error detection independent of $L$ provided physical error rates are below fixed nonzero values, and other physically reasonable assumptions. This proof is applicable to the fully fault-tolerant case only, not the case of perfect stabilizer measurements.
\end{abstract}

\maketitle

Quantum computing hardware is not expected to achieve the same level of reliability as classical computing hardware due to its complexity and reliance on the fragile phenomena of quantum mechanics. Arbitrarily reliable quantum computation can, however, be achieved through the use of quantum error correction \cite{Knil96b,Ahar97,Kita97,Fowl12e}. Bright hopes in the field of quantum error correction include the surface code and topological cluster states \cite{Brav98,Denn02,Raus07,Raus07d,Fowl09,Fowl12f}. These approaches to quantum error correction have the very experimentally reasonable requirements of a 2-D array of nearest-neighbor coupled qubits capable of implementing initialization, measurement, and one- and two-qubit unitary gates, all with error rates below approximately 1\% \cite{Wang11,Fowl11b}. Trade-offs are also possible, such as a measurement error rate of 10\% or more at the cost of somewhat lower two-qubit gate error rates \cite{Fowl12f}.

Ion traps have achieved world-leading low error single-qubit rotations \cite{Brow11}, readout \cite{Burr10}, and transport \cite{Blak11}, however single experiments designed to perform all operations have much higher error rates \cite{Hann09,Choi14}. Presently, an experimental demonstration of topological quantum error correction (TQEC) has only been possible using photons \cite{Yao12}. We are hopeful that solid-state demonstrations of TQEC shall follow shortly.

Given a 2-D nearest-neighbor coupled qubit lattice, any quantum error correction code with local stabilizers and certain additional properties can be decoded in a highly automated manner using Autotune \cite{Fowl12d}. Autotune generally requires that every isolated error leads to precisely two stabilizer measurement values changing. Errors on the qubit lattice boundaries are allowed to lead to a single stabilizer measurement value change. This is the class of quantum error correction schemes we focus on in this work. This class includes the surface code and topological cluster states. Autotune runs on a single core, an approach that would be insufficiently fast for a large quantum computer. However, a high-speed practical $O(1)$ parallel version has been proposed \cite{Fowl11b,Fowl12c}. In this work, we prove that this proposed parallel version can indeed run in the claimed $O(1)$ average time per round of error detection.

Algorithms not based on minimum weight perfect matching designed to decode topological codes do exist \cite{Harr04,Ducl09,Ducl10,Ducl13,Brav11,Sarv12,Bomb12,Woot12}, but matching approaches \cite{Brav12,Woot13} have been by far more thoroughly tested. In particular, non-matching methods have only been tested using direct, imperfect measurement of the stabilizers, while matching has been implemented with nearest neighbor faulty gates on a regular 2D lattice. Gate errors will generally only introduce short range space-time noise correlations, so it is not expected to change the problem in a fundamental way. Nonetheless, our proof removes any doubt that practical decoding of TQEC codes can be performed.

Related prior work by Sipser and Spielman \cite{Sips96} and \cite{Vide12} on the linear time decoding of low-density parity check classical codes cannot be used to prove the work considered in this manuscript. Firstly, this prior work only achieves logarithmic depth parallel processing. Secondly, nonlocal processing is used. Thirdly, these prior approaches do not guarantee correction of the theoretical maximum number of errors. Fourthly, this prior work can only be applied to expander codes, which are a particular family of classical block codes of constant rate, which the surface code is not a member of.

The discussion is organized as follows. In Section \ref{background}, a brief historical overview of minimum weight perfect matching is provided, followed by the relationship between prior work and this work. In Section \ref{mwpm}, the linear optimization problem that is minimum weight perfect matching is described. In Section \ref{smwpm}, a serial algorithm is described capable of efficiently solving this linear optimization problem. In Section \ref{serial complexity}, the average complexity of the serial algorithm when applied to problem instances associated with TQEC is shown to be $O(n)$, where $n$ is the number of detection events, defined in this same Section. The worst-case complexity is shown to be $O(n^2)$. In Section \ref{parallel complexity}, the proposed parallel implementation is described and shown to require $O(1)$ average parallel processing per round of error detection. Section \ref{discussion} concludes with a complete statement of the theorem proved in this work.

\section{Background}
\label{background}

\begin{figure}
\begin{center}
\resizebox{85mm}{!}{\includegraphics{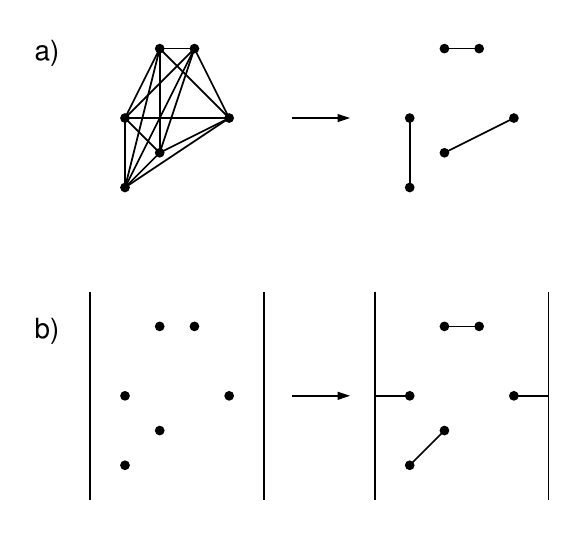}}
\end{center}
\caption{a) Standard minimum weight perfect matching input, a weighted graph, and output, a perfect matching with minimum total weight. b) Schematic minimum weight perfect matching problem associated with topological quantum error correction. Time runs vertically. Vertical lines represent boundaries. The goal is now to match each vertex with either another vertex or a boundary such that the total weight of matched edges is minimal. New vertices are constantly added to the problem.}\label{prob}
\end{figure}

Minimum weight perfect matching has a long history. The first version of the algorithm was devised by Jack Edmonds and published in 1965 \cite{Edmo65a,Edmo65b}. Conceptually, this algorithm takes a weighted graph and finds a set of edges with minimal weight sum (Fig.~\ref{prob}a). Given an even number $n$ of vertices and edges between every pair of vertices, a direct implementation of Edmonds' algorithm runs in worst-case time $O(n^4)$ \cite{Kolm09}.

A great deal of research has gone into improving the worst-case performance of Edmonds' algorithm, with performance first improved to $O(n^3)$ by Balinski \cite{Bali69}, Gabow \cite{Gabo73,Gabo76}, Kameda and Munro \cite{Kame74}, and Lawler \cite{Lawl76}. In the most recent work, this has been further improved to $O(n^{2.5})$ by Micali and Vazirani \cite{Mica80}, Gabow and Tarjan \cite{Gabo91}, and Goldberg and Karzanov \cite{Gold04}. This scaling has only been surpassed by approximate techniques presented by Duan and Pettie, which can generate a minimum weight matching within $1+\epsilon$ of optimal in $O((n/\epsilon)^2\log^3 n)$ time.

The above algorithms can cope with negative weight edges, and graphs with small numbers of edges. To the best of our knowledge, these algorithms do not currently cope with additional vertices being added during execution. Our focus will ultimately be on the graphs arising during TQEC. Such graphs involve vertices located in 3-D space-time. The separation of vertices defines the weight of a connecting edge, enabling one to omit an explicit specification of any edges at the beginning of the algorithm. Furthermore, since vertices correspond to endpoints of error chains in a quantum computer, and a quantum computer runs continuously, we must handle the case of a constant stream of additional vertices. These special properties of our problem regrettably make the existing literature difficult to use. Existing algorithms require a complete graph as input to match the error suppression performance of our algorithm, and simple construction of this complete graph would guarantee a minimum runtime of $O(n^2)$.

Given a number of vertices $n$ in a finite space-time corresponding to running a finite quantum computer for a finite amount of time, our algorithm runs on a single core in worst-case time $O(n^2)$. This is not, however, the case of greatest interest. Given an $L\times L$ qubit quantum computer running continuously with a uniform 2-D array of classical computing devices, our algorithm runs in $O(1)$ average time per round of error detection, independent of $L$, which is optimal.

\section{Minimum weight perfect matching}
\label{mwpm}

Let $G$ be a weighted graph $(V,E,W)$, meaning a set of vertices $V=\{v_i\}$, a set of edges $E=\{e_{ij}\}$ satisfying $i\neq j$ and $e_{ij}=\{v_i,v_j\}$, and a set of real weights $W=\{w_e\},e\in E$. A \emph{matching} of $G$ is a subset of edges $M\subseteq E$ such that $\forall e,f\in M,e\cap f=\emptyset$. A \emph{perfect matching} is a matching with the additional property that $\forall v\in V, \exists e\in M$ such that $v\in e$. A \emph{minimum weight perfect matching} is a perfect matching with the additional property that $\sum_{e\in M}w_e$ is minimal within the set of perfect matchings.

A \emph{complete graph} is a graph with the additional property that $\forall v_i,v_j\in V, i\neq j \Rightarrow e_{ij}\in E$. Denote the number of elements (cardinality) of a set $S$ by $|S|$. Clearly, any complete graph $G$ with an even number of vertices $|V|$ possesses a perfect matching. Let $i\in [0,\ldots,n]\subset\mathbb{Z}$. Consider $V=\{v_i\}$. We shall associate a special label with $v_0$, calling it the \emph{boundary} of $G$. Let $V^+=V-\{v_0\}$. We shall henceforth restrict ourselves to graphs with this form of index set. We shall call a matching $M$ of $G$ perfect if $\forall e,f\in M, e\cap f\neq\emptyset\Rightarrow e\cap f=\{v_0\}$ and $\forall v\in V^+, \exists e\in M$ such that $v\in e$. Note that $|V^+|$ does not need to be even for a perfect matching so defined to exist. Let $v_i,v_j,v_k$ be distinct vertices. We shall further restrict ourselves to positively weighted graphs satisfying the triangle inequality $w_{e_{ik}}\leq w_{e_{ij}} + w_{e_{jk}}$.

Let $S\subseteq V^+$. Define the \emph{hair} of $S$ to be $h(S)=\{e=\{v,w\}\in E~|~v\in S, w\in V-S\}$. In standard graph theory literature, this is typically called the boundary of $S$, however we use the term hair to avoid confusion with the boundary $v_0$ of $G$ defined above. Furthermore, the term hair gives a nice intuitive picture, as given a connected region of vertices $S$, $h(S)$ would look like the set of edges touching the surface of this region and pointing outwards. Let $\{x_e\},e\in E$ be a set of real variables. Define $O=\{S\subseteq V^+~|~|S|~{\rm odd}\}$. We impose the following conditions on $\{x_e\}$:
\begin{enumerate}
\item $\forall e\in E, x_e\geq 0$,
\item $\forall S\in O, \sum_{e\in h(S)}x_e\geq 1$.
\end{enumerate}
Let $\{y_S\}, S\in O$, be another set of real variables. We impose the following conditions on $\{y_S\}$:
\begin{enumerate}[resume]
\item $\forall S\in O, y_S\geq 0$,
\item $\forall e\in E, \sum_{S\in O~|~e\in h(S)}y_S\leq w_e$.
\end{enumerate}
Arbitrarily order the sets $O$ and $E$. Let $S_i, e_i$ denote the $i$th elements of these sets, respectively. Let $A$ denote the $|O|\times |E|$ matrix with entry $A_{ij}$ equal to 1 if $e_j\in h(S_i)$, and 0 otherwise. Let $\tilde{x}$ denote the $|E|$ entry column vector with $i$th entry $x_{e_i}$. Let $\tilde{c}$ denote the $|O|$ entry column vector with all entries 1. Let $\tilde{y}$ denote the $|O|$ entry column vector with $i$th entry $y_{S_i}$. Let $\tilde{w}$ denote the $|E|$ entry column vector with $i$th entry $w_{e_i}$. Conditions 2 and 4 can be rewritten as:
\begin{enumerate}[resume]
\item $A\tilde{x}\geq \tilde{c}$,
\item $A^T\tilde{y}\leq \tilde{w}$.
\end{enumerate}
We seek to minimize the value of $\tilde{w}^T\tilde{x}$ and maximize the value of $\tilde{c}^T\tilde{y}$.

At this point in time, some intuition into why we care about solutions of the above symmetric linear optimization problem \cite{VanR} would be of value. Consider conditions 1 and 2. Condition 1 restricts all $x_e$ variables to be positive, defining a region $P\subset \mathbb{R}^{|E|}$. Each condition $\sum_{e\in h(S)}x_e\geq 1, S\in O$ splits $\mathbb{R}^{|E|}$ in half along a plane, potentially slicing off a low $x_e$ portion of $P$. Collectively, conditions 1 and 2 define a convex subset $P'\subset P$. Given $P'$, it is clear that a well-defined, finite minimum value of $\tilde{w}^T\tilde{x}$ exists.

Let $M$ be a perfect matching of $G$ and set $x_e$ equal to 1 if $e\in M$ and 0 otherwise. Clearly, such an assignment satisfies conditions 1 and 2. Suppose it is possible to find a set of non-negative values $\{y_S\}$ such that $x_e=1$ implies $\sum_{S\in O~|~e\in h(S)}y_S = w_e$ and $\sum_{S\in O~|~e\in h(S)}y_S < w_e$ implies $x_e=0$ and the edges $e$ with $x_e=1$ form a perfect matching. Such a set $\{y_S\}$ would clearly satisfy conditions 3 and 4. Suppose in addition that $\forall S\in O, \sum_{e\in h(S)}x_e > 1$ implies $y_S=0$. We would then have $(\tilde{c} - A\tilde{x})^T\tilde{y} = 0$ and $(\tilde{w} - A^T\tilde{y})^T\tilde{x} = 0$, and hence by the complimentary slackness theorem \cite{VanR}, $\tilde{w}^T\tilde{x}=\tilde{c}^T\tilde{y}$ and $\tilde{w}^T\tilde{x}$ is minimal. Our goal, then, is to describe an efficient algorithm finding such sets of values $\{x_e\}$ and $\{y_S\}$.

\section{Serial minimum weight perfect matching}
\label{smwpm}

Start with $\tilde{x}=0$ and $\tilde{y}=0$. We shall restrict the variables $x_e$ to take the values 0 and 1. We shall call an edge $e$ with $x_e=1$ \emph{matched}, and one with $x_e=0$ \emph{unmatched}. We shall ensure at all times that the set of matched edges is a matching. Given a matched edge $e$, vertices $v,w\in e$ shall also be called \emph{matched} with the exception of the boundary vertex $v_0$, which shall always be called \emph{unmatched} regardless of how many matched edges it belongs to. Any vertex not belonging to a matched edge shall also be called \emph{unmatched}. With the stated initial variable assignments, all vertices are initially unmatched.

\begin{figure}
\begin{center}
\resizebox{85mm}{!}{\includegraphics{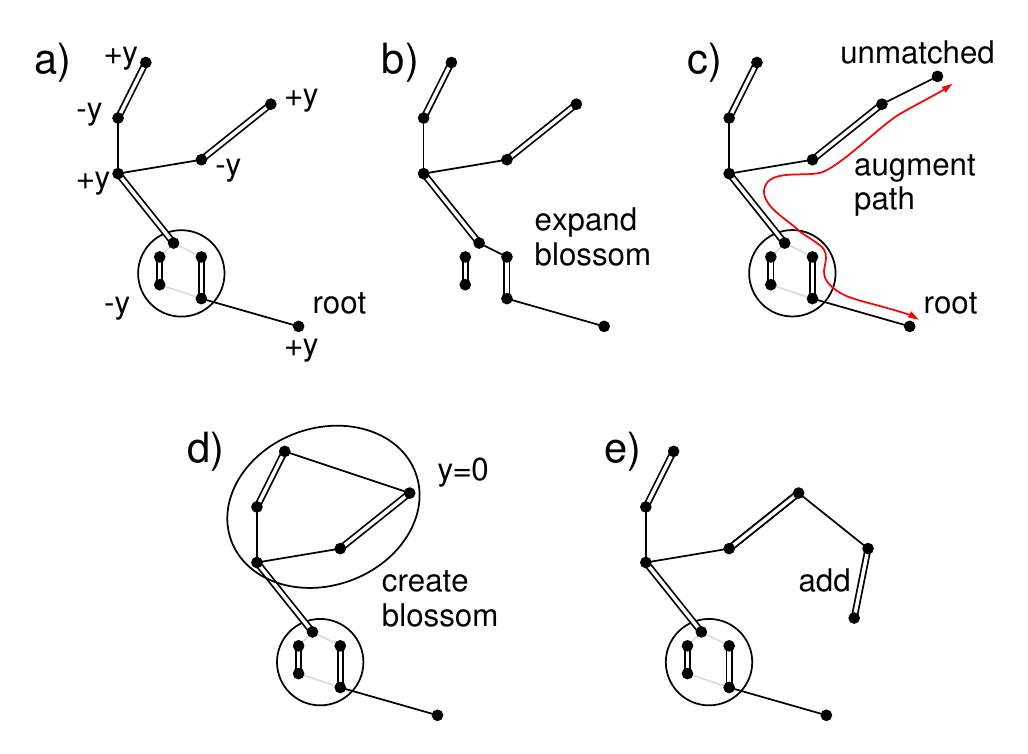}}
\end{center}
\caption{All required alternating tree manipulations.
a)~Increase outer node and decrease inner node $y$ values, maintaining the tightness of all tree edges and potentially creating new tight edges connected to at least one outer node. b)~Inner blossoms
with $y=0$ can be expanded into multiple inner and outer nodes and potentially some nodes that are no longer part of the tree. c)~Outer--matched tight edges can be used to grow the alternating tree.
d)~Outer--inner tight edges can be ignored. e)~Outer--outer tight edges
make cycles that can be used to make blossoms. f)~When another unmatched vertex $v$ is found, the path from the unmatched vertex within the root node through the alternating tree to $v$ is augmented, meaning matched edges become unmatched and unmatched edges become matched. This strictly increases the total number of matched vertices.}\label{Tree}
\end{figure}

Define an edge $e$ satisfying $\sum_{S\in O~|~e\in h(S)}y_S = w_e$ to be \emph{tight}. An edge that is not tight is called \emph{slack}. We shall ensure that all matched edges are tight, but not all tight edges will be matched. Define a \emph{node} to be a vertex or blossom, where a \emph{blossom} is an odd cycle of nodes constructed as described in step (g) below, an example of which is shown in Fig.~\ref{Tree}d. Define a blossom to be
\emph{unmatched} if it contains an unmatched vertex. An \emph{alternating tree} is
a tree of nodes rooted on an unmatched node such that every path
from the root to a leaf consists of alternating unmatched and
matched tight edges. Alternating trees can by this definition only branch from the root and
every second node from the root. Define branching nodes to be \emph{outer}.
Define all other nodes in the alternating tree to be \emph{inner}. Fig.~\ref{Tree} shows all necessary alternating tree manipulations.

A number of invariants are maintained during the execution of the algorithm. Many have already been mentioned, however for convenience we gather them all here.
\begin{enumerate}[resume]
\item $\forall e\in E, x_e\in\{0,1\}$
\item $\forall S\in O, y_S\geq 0$
\item $\forall e\in E, \sum_{S\in O~|~e\in h(S)}y_S\leq w_e$
\item $\{e\in E~|~x_e=1\}$ is a matching
\item $x_e=1\Rightarrow \sum_{S\in O~|~e\in h(S)}y_S=w_e$
\item $\sum_{S\in O~|~e\in h(S)}y_S<w_e\Rightarrow x_e=0$
\item $v$ unmatched and not in an alternating tree implies $y_{\{v\}}=0$
\end{enumerate}
Note that while conditions 1, 3, and 4 are implied by conditions 7, 8, and 9, condition 2 will only be satisfied when the algorithm terminates with all vertices matched.

Define a \emph{growth edge} to be a tight edge connecting an outer node to anything other than an inner node. Given a weighted graph $G$, the following algorithm finds a minimum
weight perfect matching.
\begin{enumerate}[label={\bf (\alph*)}]
\item If there are no unmatched vertices in $V^+$, return the set of matched edges.

\item Choose an unmatched vertex $v\in V^+$ to be the root of an alternating tree.

\item If there are no growth edges, increase the value of $y$ associated with each outer node while simultaneously decreasing the value of $y$ associated with each inner node until a growth edge is created, or an inner blossom node $y$ variable becomes 0 (Fig.~\ref{Tree}a).

\item If an inner blossom node $y$ variable becomes 0 and there are no growth edges, expand that blossom and return to step (c) (Fig.~\ref{Tree}b).

\item Choose a growth edge $e$.

\item If $e$ leads to an unmatched node, or a node matched to the boundary (which is itself an unmatched node), augment the path (unmatched$\leftrightarrow$matched) from the unmatched vertex within the root node to the unmatched vertex within the found unmatched node (Fig.~\ref{Tree}c). Destroy the alternating tree, keeping any newly formed blossoms. Return to step (a).

\item If $e$ leads to an outer node, add the growth edge to the alternating tree. There will now be a cycle $C\subset V^+$ of odd cardinality $|C|$. Collapse this cycle into a new blossom and associate a new variable $y_C=0$ (Fig.~\ref{Tree}d). Return to step (c).

\item Add the growth edge and the matched edge leading from the growth edge to the alternating tree (Fig.~\ref{Tree}e). Return to step (c).
\end{enumerate}

\section{Serial minimum weight perfect matching complexity}
\label{serial complexity}

\begin{figure}
\resizebox{60mm}{!}{\includegraphics{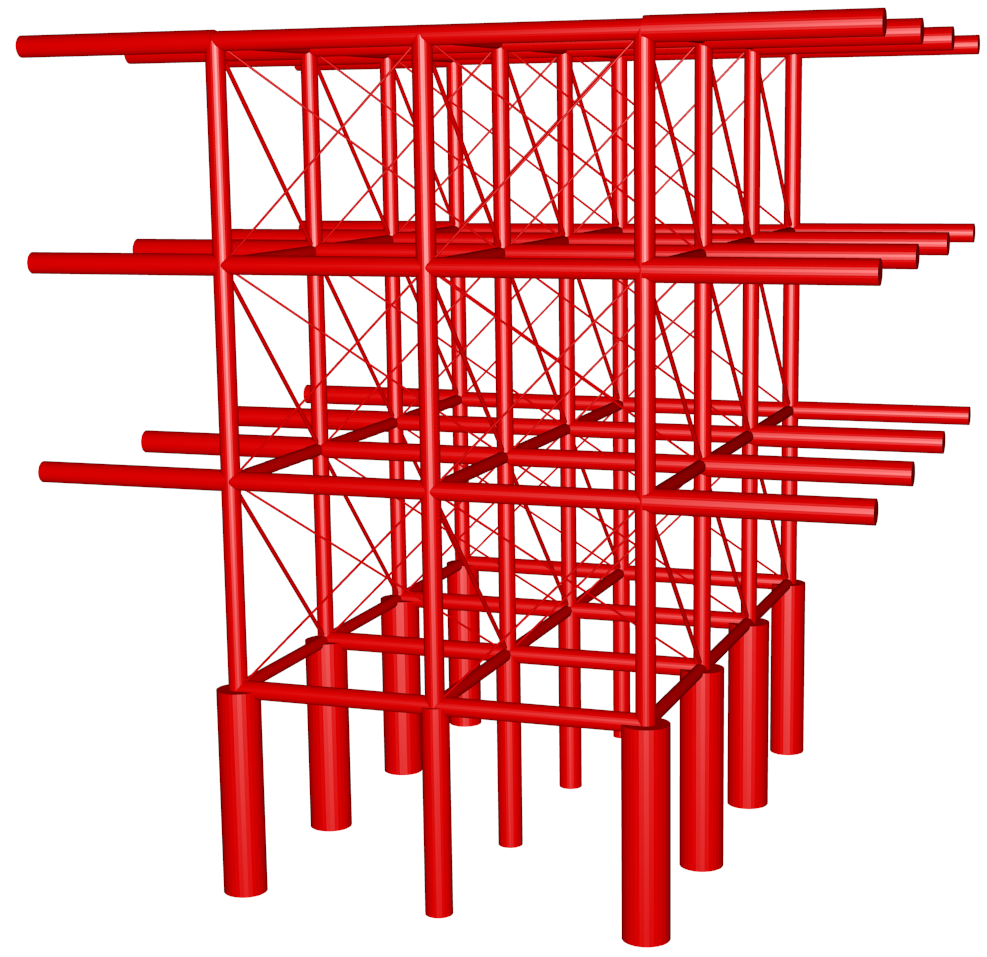}}
\caption{Autotune generated (primal) nest of a distance 4 surface code with depolarizing noise. \emph{Primal nests} are associated with $Z$ error detection. Cylinders (sticks) do not have equal diameter, accurately representing the diverse range of probabilities of various pairs of detection events. Note also the many diagonal sticks, which are associated with errors that propagate to space-time location separated by more than one unit of space and/or time.}
\label{Autotune_sc}
\end{figure}

The algorithm described in the previous Section is quite general, however we are only interested in the complexity of minimum weight perfect matching on graphs generated during TQEC. We shall be using the concepts of a \emph{nest} and a \emph{detection event}, terminology introduced in \cite{Fowl12d}. A \emph{nest} is a 3-D structure of cylinders (\emph{sticks}), each of whose diameter is proportional to the total probability of detection events at the endpoints of the sticks arising from single errors. A \emph{detection event} is simply a local pattern of measurements indicating the nearby presence of an error. For convenience of discussion, we say that a \emph{ball} is located at the points where stick endpoints meet. Fig.~\ref{Autotune_sc} contains an example of a nest associated with $Z$ error detection in a distance 4 surface code. The terminology balls and sticks conveniently distinguishes nests from graphs which contain vertices and edges.

The weight of a stick with probability $p$ is defined to be $-\ln p$. In a running quantum computer, detection events are observed at random locations. Each detection event is associated with a vertex in a graph. The weight of an edge connecting a pair of vertices or a vertex to a boundary is defined to be the minimum weight connecting path through the nest. With this definition, edges do not need to be explicitly constructed and the implicitly defined edges of the graph automatically satisfy the triangle inequality. Generating a nest with $n$ vertices can be completed in $O(n)$ time. The input to our algorithm can thus be generated optimally.

\begin{figure}
\begin{center}
\resizebox{85mm}{!}{\includegraphics{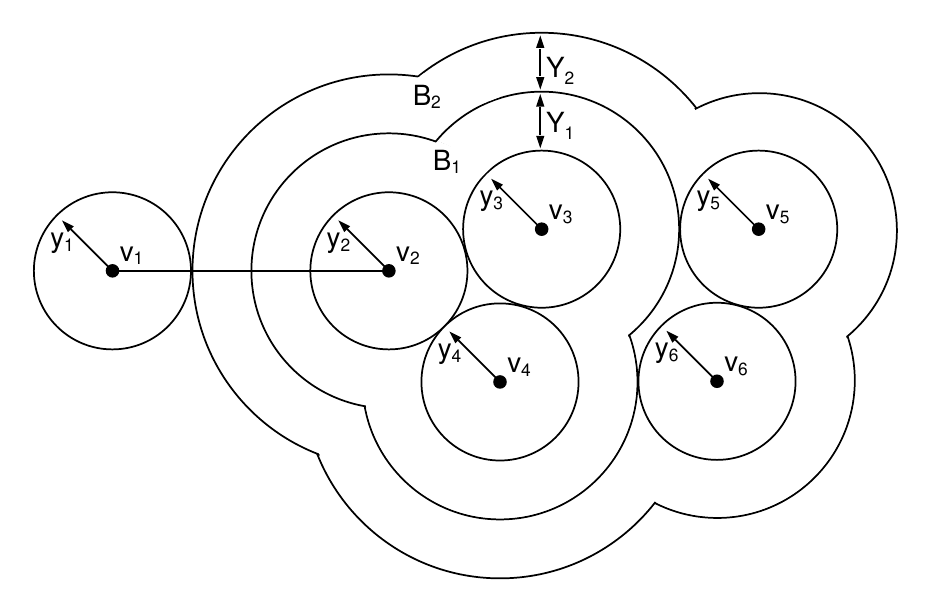}}
\end{center}
\caption{An example of a tight edge. $Y_1=y_{\{v_2,v_3,v_4\}}$. $Y_2=y_{\{v_2,v_3,v_4,v_5,v_6\}}$. $y_i=y_{\{v_i\}}$. Edge $e_{12}$ has
the property that $w_{e_{12}}-y_1-y_2-Y_1-Y_2=0$. Abbreviating $e_{12}$ to $e$, this can also be expressed as $\sum_{S\in O~|~e\in h(S)}y_S=w_e$.}\label{tight_edge}
\end{figure}

The variables $y_S$ in a graph associated with a nest can be visualized as exploratory regions. An example of a tight edge with various $y_S$ variables visualized in this manner can be found in Fig.~\ref{tight_edge}. The utility of this visualization lies in the realization that tight edges can be detected by keeping track of when expanding exploratory regions collide with vertices, boundaries, or other exploratory regions. This enables the algorithm described in the previous Section to only generate explicit edges as they are required. We now need to determine the probability distribution of the number of operations $n_{\rm op}$ required to successfully match a vertex to another vertex or the boundary.

It will be necessary for any sufficiently large quantum computing system to be built in a modular manner, rather than manufactured in one enormous piece. Modularity has a number of advantages in addition to enabling practical manufacturing --- one can set reasonable manufacturing standards and discard modules that do not meet them. This means one can control both the density and distribution of hardware faults, meaning qubits and couplers that are non-functional at time of manufacture. We do not require that all components in a given module work, however we do require that modules be assembled in a manner that sets a strict upper bound on the size of any patch of connected non-functional hardware. Patches of non-functional hardware necessitate measuring larger stabilizers encircling these patches \cite{Stac09}. Larger stabilizers cannot be measured as reliably as small stabilizers. If the quantum computer construction were not controlled to set a reasonable upper bound on the stabilizer size, stabilizers of arbitrary size would need to be measured, and above a certain size the stabilizer measurement results would essentially be random, leading to a breakdown of the error correction procedure. Any density of excessively large stabilizers would limit the maximum reliability of the quantum computer.

Define the \emph{degree} $b$ of a ball to be the number of sticks ending there. By setting a maximum stabilizer size at time of manufacturing, we set a maximum \emph{degree} $b_{\rm max}$. We assume that if it is possible for the qubit state to leak to a non-computational state or for the qubit be lost entirely, the underlying TQEC code can detect these errors. Topological cluster states \cite{Raus07,Raus07d,Fowl09} provide an example of a topological code capable of detecting leakage \cite{Ghos13,Fowl13f} and loss \cite{Barr10}.

An isolated gate error leads to detection events at the endpoints of a particular stick. Leakage or loss leads to the need to measure a larger stabilizer encompassing the connected region of leaked or lost qubits \cite{Barr10,Herr10}. A single leakage or loss of event effectively leads to the merging of a pair of neighboring balls, which can equally well be visualized as the labeling of the stick connecting these balls. Gate errors, qubit state leakage, and qubit loss can therefore all be thought of as highlighting a particular stick. We need to ensure that connected regions of highlighted sticks are small on average. A connected region of highlighted sticks connecting boundaries within a topological code or encircling some structure in a topologically non-trivial manner results in an undetectable error.

Fortunately, it is known that provided an infinite extent graph has a finite maximum degree $b_{\rm max}$ and the probability of highlighting any edge is below some fixed nonzero value (the percolation threshold), the probability of any given edge belonging to a connected region of highlighted edges of size $n$ decays exponentially with $n$ \cite{Mens86}. In the case of a TQEC nest associated with a quality controlled modular quantum computer, a percolation threshold exists as there is a maximum stabilizer size and therefore errors cannot accumulate for long before being detected, meaning a finite maximum stick probability $p_{\rm max}<1$. To state all of this in a manner convenient for our needs, the probability of an error chain of length $L$ sticks beginning on any given vertex (detection event) can be upperbounded by $Ax^L$ for some $0<x<1$ and some $A$ such that $Ax < 1$.

A finite $p_{\rm max}<1$ directly implies a nonzero minimum stick weight $w_{\rm min}$. Under the assumption that quantum computer errors occur with some nonzero minimum probability, there will also be a nonzero $p_{\rm min}$, and hence a finite $w_{\rm max}$. Let $R=\lceil w_{\rm max}/w_{\rm min}\rceil$.

Consider an error chain $v_1v_2$. We wish to calculate an upper bound $n_{\rm av}$ on the average number of other error chains $u_1u_2$ sufficiently nearby to enable an alternating tree to grow. If the $v_1v_2$ chain has length $L_v$, made up of sticks of weight $w_{\rm max}$, and the $u_1u_2$ chain has length $L_u$, also made up of sticks of weight $w_{\rm max}$, then provided some $u_j$ is within $R(L_v+L_u)$ sticks of some $v_i$, there is a chance of a tight edge $v_iu_j$. The number of balls reachable by any path of $R(L_v+L_u)$ sticks from any given ball is at most $b_{\rm max}^{R(L_v+L_u)}$. Focusing temporarily on just one vertex $v_i$, given an error chain ends here, the fraction of error chains containing $L_v$ sticks is at most $(1-x)x^{L_v-1}$. The average number of other error chains $u_1u_2$ sufficiently nearby either $v_1$ or $v_2$ to enable an alternating tree to grow is therefore no more than
\begin{equation}
2\sum_{L_v=1}^\infty(1-x)x^{L_v-1}\sum_{L_u=1}^\infty b_{\rm max}^{R(L_v+L_u)}Ax^{L_u},
\end{equation}
which simplifies to
\begin{equation}
n_{\rm av}=\frac{A(1-x)xb_{\rm max}^{2R}}{(1-xb_{\rm max}^R)^2}.
\end{equation}

The value of $x$ depends on physical error rates and can, therefore, in principle be made arbitrarily low. We restrict ourselves to hardware with sufficiently low physical error rates to ensure $n_{\rm av}<1$. For the surface code, $R=2$ and $b_{\rm max}=12$, and hence the above suggests $x\sim 10^{-5}$. Our simulation results in \cite{Fowl12c} suggest $x>10^{-3}$.

It is highly likely, although not yet proven, that local behavior is maintained all the way up to the threshold error rate. This makes theoretical sense as only at the threshold error rate and above are error patterns ambiguous on an infinite scale leading to arbitrarily large incorrectly identified error patterns. Locality should prevail at any error rate below threshold. Our low proven value of $x$ should be clearly understood to be the result of the approximations and loose bounds used in the proof, rather than being of fundamental nature.

In the $n_{\rm av}<1$ regime, large alternating trees are exponentially unlikely to grow. It should be clearly understood that this means that a low density of detection events at the ends of short error chains is on average locally matchable. Global information is not required, and indeed the matching problem will decompose into small local clusters of detection events that are algorithmically forbidden from interacting. The detection events in each cluster can only be matched amongst themselves.

Given a finite-size quantum computer running for a finite amount of time resulting in $n$ detection events, minimum weight perfect matching any given vertex results in the systematic one-way exploration of at most the entire finite volume, which takes $O(n)$ time given the constant density of vertices. This means finite-volume TQEC graphs can be matched in worst-case $O(n^2)$ time.

The run-time complexity of minimum weight perfect matching a small cluster of $n$ detection events is again at worst $O(n^2)$. Combining this with the exponential distribution of cluster sizes and consequent constant average cluster size independent of problem size leads to an average runtime to match a single detection event of $O(1)$. Given $n$ detection events, the input structure and output can both therefore be generated in $O(n)$ time. This has been empirically corroborated in \cite{Fowl12c}, where the decoding time per round of error detection was observed to grow in proportion to the area of the surface code considered.

\section{Parallel minimum weight perfect matching complexity}
\label{parallel complexity}

Consider an $L\times L$ 2-D array of qubits with some constant density of associated classical processing elements, each nominally servicing $N$ qubits, although note in practice that these processing elements can communicate, and with low probability matching a single detection may involve communicating with a large region. The 2-D array of classical processing elements is assumed to each have some large but finite amount of local memory and the capacity to communicate with the eight neighboring processing elements. Each processing element would be an ASIC (Application-Specific Integrated Circuit) custom designed to run minimum weight perfect matching only.

Each ASIC would nominally be responsible for some square patch of $N$ qubits total. The qubits in this patch would generate a random stream of detection events. For convenience, each ASIC would also receive notification of detection events in the eight neighboring square patches.

Initially, consider just one ASIC working without communicating with its neighbors. If a given detection event can be matched to some other detection event in this ASIC's local patch without creating an alternating tree with exploratory regions that bleed into the neighboring patches, the ASIC would be permitted to proceed with the matching without notifying its neighbors.

If, however, an alternating tree not exclusively within the local patch is required, communication is necessary. If the alternating tree does not bleed outside the neighboring eight patches, the ASIC can proceed and simply notify each neighboring ASIC that had its patch touched of the details performed in that region. If these modifications are inconsistent with what has already been done there, all detection events involved in this inconsistency would be unmatched and made the responsibility of the middle ASIC. All other ASICs involved in the inconsistency could simply be stalled while this occurs.

If an alternating tree needs to span many patches, an arbitrary ASIC can be chosen responsible for the entire tree, all other ASICs with detection events associated with this alternating tree in their local patch can be stalled, and the single chosen ASIC can proceed as normal, requesting data from and writing data to nonlocal patches through sequential nearest neighbor communication.

Another possibility is an alternating tree extending further in the past than is stored in local memory. While one would choose a sufficiently large local memory to make this unlikely, it is not a possibility that can be eliminated. To handle this, we must restrict our interest to quantum computations of finite duration, a reasonable assumption given we are unlikely to want to run a quantum algorithm for more than a year, and have slower external storage of all of the previous detection events and all matching data no longer stored in local memory. If we ensure that the probability of requiring external data is sufficiently low, the impact of accessing external data on the average detection event matching time can be made negligible.

Clearly, large alternating trees will take longer to process, however nonlocal communication adds at worst polynomial overhead to a procedure that runs in at worst $O(n^4)$ time, using the original unoptimized Edmonds' algorithm. Given larger alternating trees are exponentially unlikely, the average time spent matching a given detection event is still a well-defined constant value. This value takes into account the possibility of being stalled while some other ASIC uses local memory.

The average number of detection events per local patch per round of error detection is also a well-defined constant value independent of problem size. One can therefore define an average required classical processing time $T_c$ per round of error detection, a time which includes a certain amount of probabilistic stalling.

Define $T_q$ to be the time required to perform a round of error detection using the quantum hardware assuming any probabilistic execution paths succeed. For example, if some ancilla state is required, and this state is probabilistically prepared, define $T_q$ to be the time required to perform error detection assuming this preparation succeeds the first time. Assume furthermore that $T_q$ is defined with reference to a region of quantum hardware with no non-functional components. The purpose of $T_q$ is simply to provide a well-defined heartbeat for the quantum computer.

We assume it is possible to build sufficiently fast ASICs such that $T_q=2T_c$. This means that, in addition to any stalling imposed by large alternating trees, which is already included in $T_c$, on average every ASIC will be idle by choice a significant fraction of the time. This fraction will be less than 50\%, as we shall see.

On average, each ASIC has plenty of time to cope with its stream of detection events and communicate results with its neighbors. However, with exponentially small probability, and arbitrarily large alternating tree can be required which delays all of the ASICs it touches. Note crucially, however, that ASICs not touched by any large alternating tree will continue to process as normal without delay. When the problematic large alternating tree is finally cleared, and the difficulty of matching detection events trends back to average difficulty, the fact that we have designed the ASICs to run sufficiently fast that $T_q=2T_c$ means they will be able to catch up. The parallel algorithm is thus asynchronous, however any given ASIC will fall linearly behind the average progress mark only with exponentially small probability, and with no global impact. The alternating delay and then catch up cycle reduces the idle time below 50\%.

It is reasonable to assume that any quantum computation must take at least $O(\log L)$ time since this is the minimum number of rounds of error detection required to implement even a single layer of robust fault-tolerant logical gates. Finishing off the classical processing at the end of the algorithm also takes $O(\log L)$ time on average due to exponentially unlikely hard matching instances. In more detail, the volume of the entire algorithm is $O(L^2\log L)$. The probability of considering information a distance $r$ away from any given initial vertex is $O(e^{-ar})$ for some positive constant $a$, so the average maximum value of $r$ only grows logarithmically with the number of matchings. We can therefore upper bound the average maximum radius $r_{\rm max}$ by a value $O(\log(L^2\log L))=O(\log L)$. The average processing time per round of error detection is therefore a constant independent of $L$.

\section{Conclusion}
\label{discussion}

We have proved that, given the following ingredients:
\begin{enumerate}
\item an $L\times L$ qubit quantum computer
\item a modular architecture such that there is a finite maximum number of non-functioning qubits in any given connected defective patch
\item gate, leakage and loss error rates below some set of nonzero values
\item a uniform 2-D array of finite speed processing elements with finite local memory and the ability to communicate with their nearest neighbors at finite speed
\item external memory with capacity sufficient to store all detection events and matching data for the duration of a temporally finite quantum algorithm
\end{enumerate}
it is possible to solve the minimum weight perfect matching problem in a globally optimal manner with $O(1)$ average cost per round of error detection independent of $L$.

\section{Acknowledgements}
\label{ack}

We thank David Poulin for constructive comments on this work. This research was conducted by the Australian Research Council Centre of Excellence for Quantum Computation and Communication Technology (project number CE110001027), with support from the US National Security Agency and the US Army Research Office under contract number W911NF-13-1-0024. Supported by the Intelligence Advanced Research Projects Activity (IARPA) via Department of Interior National Business Center contract number D11PC20166.  The U.S. Government is authorized to reproduce and distribute reprints for Governmental purposes notwithstanding any copyright annotation thereon.  Disclaimer: The views and conclusions contained herein are those of the authors and should not be interpreted as necessarily representing the official policies or endorsements, either expressed or implied, of IARPA, DoI/NBC, or the U.S. Government.

\bibliography{../References}

\end{document}